\documentclass[]{aastex631}

\usepackage{tikz}
\usepackage{float}
\usepackage{CJK}
\usepackage{multirow}
\let\splitbox\relax
\usepackage{adjustbox}

\usepackage{gensymb}
\usepackage{enumitem} 
\usepackage{amsmath,amssymb,amstext}
\usepackage{graphicx}
\usepackage{hyperref}
\usepackage[utf8]{inputenc}
\usepackage{soul}

\begin{document}

\author{D. N. Garzon}
  \affiliation{Max-Planck-Institut für Astronomie, Königstuhl 17, 69117 Heidelberg, Germany}
\affiliation{Illinois Center for Advanced Studies of the Universe,
University of Illinois at Urbana-Champaign, Urbana, IL 61801, USA}
 \affiliation{Yachay Tech University, 100119-Urcuqui, Ecuador}

\author{Neige Frankel}
 \affiliation{Canadian Institute for Theoretical Astrophysics, University of Toronto, 60 St. George Street, Toronto, ON M5S 3H8, Canada}
  \affiliation{Department of Astronomy and Astrophysics, University of Toronto, 50 St. George Street, Toronto, ON M5S 3H4, Canada}
\author{Eleonora Zari}
 \affiliation{Max-Planck-Institut für Astronomie, Königstuhl 17, 69117 Heidelberg, Germany}
\author{Maosheng Xiang}
\affiliation{National Astronomical Observatories, Chinese Academy of Sciences, Beijing 100012, China}
\affiliation{Institute for Frontiers in Astronomy and Astrophysics, Beijing Normal University, Beijing 102206, China}
\author{Hans-Walter Rix}
 \affiliation{Max-Planck-Institut für Astronomie, Königstuhl 17, 69117 Heidelberg, Germany}

\date{\today}




\title{The Age-Dependent Vertical Actions of Young Stars in the Galaxy}

%
%
%
%
%

%
\begin{abstract}

Stars in the Galactic disk are born on cold, nearly circular orbits with small vertical excursions. After their birth, their orbits evolve, driven by small- or large-scale perturbations in the Galactic disk's gravitational potential. Here, we study the vertical motions of young stars over their first few orbital periods, using a sample of OBA stars from \textit{Gaia} E/DR3, which includes radial velocities and ages $\tau$ from LAMOST. We constructed a parametric model for the time evolution of the stellar orbits' mean vertical actions $J_z$ as a function of Galactocentric radius, $R_{\mathrm{GC}}$. 

Accounting for data uncertainties, we use Markov Chain Monte Carlo (MCMC) analysis in annuli of Galactocentric radius to constrain the model parameters. Our best-fit model shows a remarkably linear increase of vertical actions with age across all Galactocentric radii examined. Orbital \textit{heating} by random scattering could offer a straightforward interpretation for this trend.
However, various other dynamical aspects of the Galactic disk, such as stars being born in a warped disk, might offer alternative explanations that could be tested in the future. 

\end{abstract}

%
%
%
%
%
%

\section{Introduction}\label{sec:intro}


Disk galaxies, including the Milky Way, are thought to build their stellar mass mostly via star formation, i.e. either from their own gas or from gas brought in by mergers \citep[e.g.][]{Tinsley87,FallEfstathiou,Steinmetz95}. The dynamical evolution and spatially varying star formation histories shape these galaxies after initial mass accumulation.  At the late stages of galaxy evolution, stars are thought to form on nearly circular orbits from thin disks of gas. The time-varying and non-axisymmetric structures present within disks can subsequently heat stellar orbits by increasing their radial actions (or eccentricities) and their vertical actions (or random motions in the vertical direction) \citep{Spitzer53}. These processes have been observed in simulations \citep[e.g.][]{sellwood_2014, JenkinsBinney} and have evidence in external galaxies \citep[e.g.][]{courtois2015giant, vasiliev2022radialization, vera2016conservation}.

Mechanisms like minor mergers and external perturbations can abruptly increase stars' vertical motions, whereas others, such as interactions with giant molecular clouds (GMCs) or other massive structures, contribute to a more gradual heating effect \citep{qu2011minor, spitzer_1951, Lacey1984, carlberg1987vertical, jenkins1990spiral}. In addition, vertical breathing waves induced by the bar and spiral arms \citep{barbanis1967orbits, masset_1997, monari2015vertical, Grand2016} may cause heating.

In the Milky Way, star-by-star astrometric and spectroscopic data open a window to a detailed picture of disk galaxy evolution. This data reveals that both stars and the molecular gas from which they form exhibit low velocity dispersion initially \citep[e.g.][]{ bovy2012vertical,li2013500}. In the solar neighborhood,  older stars present higher vertical velocity dispersion ($ \sigma_z$) than younger stars \citep[e.g.][]{ hayden2020galah,yu2018age}. As mentioned before, different dynamical processes can induce heating, i.e., acquisition of vertical motion by stars over time.  
The secular heating of stars has been studied through age-velocity dispersion relations \citep{wielen_1977, quillen2001saturation, Aumer2009, aumer_2016, sanders_das_2018}, suggesting that the present-day vertical-velocity dispersion continuously rises and scales as $\sigma_z \sim \tau^{0.5}$, where $\tau$ represents stellar ages \citep{nordstrom_etal_2004, holmberg2009geneva}. The present-day distribution of vertical motions of stars has also been modeled as a combination of their birth velocity dispersion, followed by a heating period that includes radial dependence \citep{ting_rix_2019}.

 In recent years, the \textit{Gaia} collaboration has released the most extensive census of positions, velocities, and other stellar parameters of billions of stars in the Milky Way, including luminous, massive, and hot stars. 

These stars are  massive, young, and can be used to study the spiral structure and the dynamics of the young Milky Way disk \citep{xu2018spiral, Zari2021, Poggio2021}. 

Stars' orbits at birth can yield insights into the dynamic state of the gas from which they formed. Subsequent vertical evolution may provide details regarding the heating mechanisms in the disk. However, the characterization of stellar heating on short timescales remains an open question. Addressing these questions involves studying young stars in the Milky Way's disk. The relationship between the vertical velocity dispersion ($\sigma_z$) and stellar age in the galaxy has been a subject of debate primarily due to the challenges in determining stellar ages and the varying selection criteria across different surveys. However, precise parallax data from the Gaia mission \citep{gaia_prusti_2016, Gaia_eDR3_Brown_2021}, combined with robust spectroscopic stellar parameters delivered from large spectroscopic surveys, yields uniform age estimates precise to the $\sim10$\% level for a large number of stars \citep[e.g.][]{xiang_rix_2022}, enabling new insights into the vertical evolution history of the Milky Way. Understanding the age-vertical action relation is a crucial step in understanding the evolution of the galaxy. 

In this work, we use the positions and velocities of young OBA stars from \textit{Gaia} EDR3 as presented by \citet{Zari2021} to map the present-day $J_z$ as a function of stellar age $\tau$ and the current Galactocentric radius. This approach offers new insights into the dynamical state of the disk. We use $J_z$ because $\sigma_z$ can change due to both gradual variations in the mid-plane baryon density and radial migration. Therefore, to better characterize the global changes in stellar orbits, it is preferable to describe the vertical motion of stars by their vertical actions and to understand heating as an increase in $J_z$ \citep{ting_rix_2019}.

The paper is organized as follows. Section \ref{sec:sample} introduces the OBA star sample and its age distribution. Section \ref{sec:Vertical Heating Model } describes the theoretical model for the vertical action with respect to galactocentric radii and the subsequent Bayesian inference. In Section \ref{sec:Results}, we present the  posteriors of the model and we expand the  results that show a linear behavior of young stars' vertical action with age. In section \ref{sec:Discussion and Summary} we briefly discuss the robustness of the results, data selection, and dust extinction and possible physical explanation.  
    %

%
%
%
%
%
%
\section{Data: Luminous Hot Star  Sample }\label{sec:sample}

\begin{figure}[h!]
    \centering
    \includegraphics[width=0.95\columnwidth]{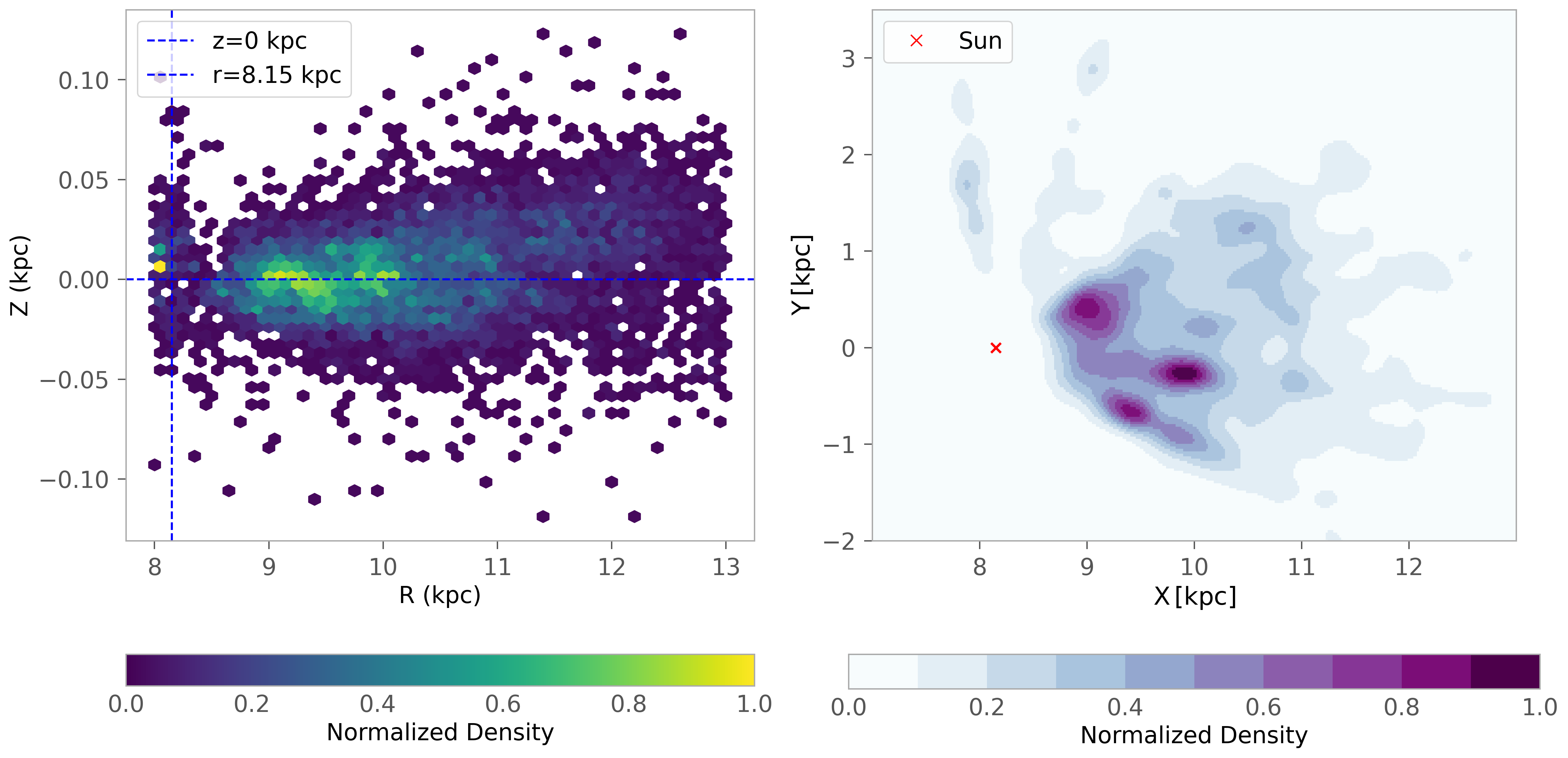}
    \caption{\textbf{Left:} Spatial distribution of the OBA stars in our filtered dataset (Z21-500) projected onto the Galactic $R-Z$ plane. The data range spans from $R = 8  $kpc to $R = 13  $kpc for the Galactocentric radius. The plot illustrates the density of stars within this region using a hex-bin plot, with the density scale shown at the bottom. \textbf{Right:} Spatial distribution of star sample projected onto the  $X-Y$ plane. The Galactic center is at \((X,Y)=(0,0)\) and the Solar position is illustrated with the red cross.  The figure shows the kernel density estimate (KDE). The colorbar indicates the density normalized with respect to the maximum value of the KDE. }
    \label{fig:combined_distribution}
\end{figure}

Our objective is to investigate the vertical actions of stars at their birth within the Milky Way disk. Therefore, we require a sample with estimable ages of young stars, specifically, those younger than a few dynamical periods ($\lesssim 0.5$~Gyr). Additionally, this sample should cover a substantial portion of the Galactic disk and possess complete 6D phase space information, which is essential for accurately estimating the orbital actions.

We constructed our sample based on the catalog of OBA star candidates (Z21) presented by \cite{Zari2021}. They used a combination of \textit{Gaia} EDR3 photometry and astrometry, along with 2MASS photometry, to select candidate OBA stars and exclude contamination from red giant branch and asymptotic giant branch stars. For their filtered sample, they further selected stars in proximity to the mid-plane ($|z| < 300$ pc) and belonging to a low-velocity dispersion component in the vertical direction \citep[see sections 2 and 4 of][]{Zari2021}.

\begin{figure*}
\centering
\begin{adjustbox}{width=\textwidth, trim={0.5 0.0 0.5 0.0}, clip}
\includegraphics{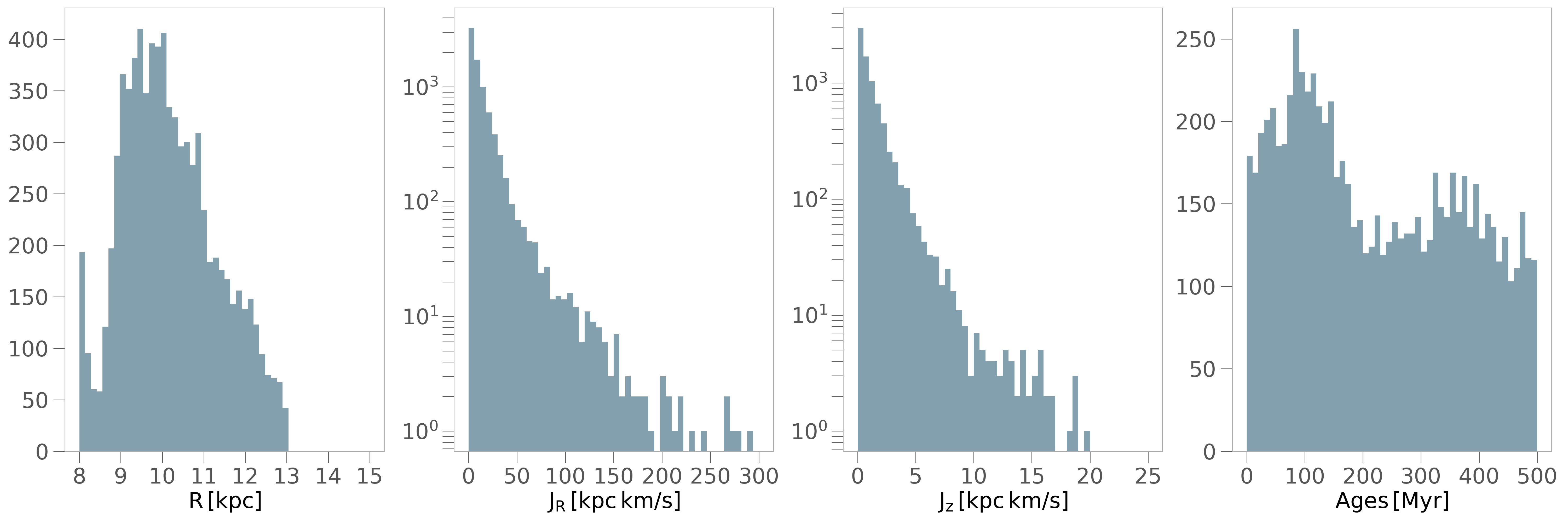}
\end{adjustbox}
\caption{1-D Distribution of stars used in our study. It is shown the Galactocentric radius \( R \), radial action \( J_R \), vertical action \( J_z \), and ages \( \tau \). Note second and third panel, a base-10 log scale is used for the $Y$ axis.}
\label{fig/hst_ages.png}
\end{figure*}

We then cross-matched our data set with the LAMOST catalog presented in \cite{xiang_etal_2022} to obtain stellar parameters, including the radial velocity \( v_\mathrm{los} \) (see Figures \ref{fig:combined_distribution} and \ref{fig/hst_ages.png}). The values for \( v_\mathrm{los} \) in \cite{xiang_etal_2022} are derived from the redshift estimates of the LAMOST spectra. However, the realistic \( v_\mathrm{los} \) uncertainty is primarily due to wavelength calibration error, given the low spectral resolution of LAMOST. As a result, the uncertainty of radial velocity reported in \cite{xiang_etal_2022} is likely underestimated. In their comparison of velocity measurements across multiple visits for common stars, \cite{xiang_etal_2022} found a typical scatter in \( v_\mathrm{los} \) of about 10 km/s. We adopt this value for the noise model of \( v_\mathrm{los} \) for all stars in our study. The observed scatter from repeated measurements also includes contributions from binary stars. Since binary stars are not directly modeled in this work, we account for their effects on the velocity by marginalizing over the observed broadening of the velocity dispersion attributable to binary star populations.

We estimate ages by employing a Bayesian isochrone fitting procedure as presented in \citet{xiang_rix_2022}. The inputs for our method include parallax, the LAMOST spectroscopic stellar parameters ($T_{\mathrm{eff}}$, $\log g$, [Fe/H]), and the multi-band photometry from Gaia ($G$, $BP$, $RP$) and 2MASS ($J$, $H$, $K$). For age inference, we adopt MIST stellar isochrones \citep{dotter_etal_2016, 2016choi}. To ensure the high precision of the spectroscopic stellar parameters, and thus the accuracy of the age estimates, stars with a LAMOST spectral S/N lower than 30 are excluded from our sample.

Using the astrometric parameters ($l, b, \mu_{l}, \mu_{b}$), distances, and radial velocities, we compute the six-dimensional phase-space coordinates in the cylindrical Galactocentric reference frame ($R, \phi, Z, V_R, V_\phi, V_Z$) and the subsequent vertical actions. We adopt a left-handed coordinate system for the frame transformations, in which the x-axis is directed towards the Galactic center, and the y-axis increases to the right when viewed from the North Galactic Pole, opposite the direction of Galactic rotation. For the frame transformations, we adopt a vertical distance of the Sun above the plane of $Z_\odot= 20.8$ pc \citep{bennett_bovy_2019}, a distance of the Sun to the Galactic center of $R_\odot = 8.15$ kpc \citep{reid2019trigonometric}, and a circular velocity at the Sun's radius of $V_c(R_\odot) = 236$$\,\mathrm{km\,s}^{-1}$  \citep{reid2019trigonometric}.
We assume a peculiar velocity of the Sun with respect to the local standard of rest of $(U_{\odot}, V_{\odot}, W_{\odot}) = (11.1, 11, 7.25)$ km s$^{-1}$ \citep{schoenrich_2010, reid2019trigonometric}.


For distances smaller than $\sim 4$ kpc, the dominant contributors to the fractional error distribution are parallax uncertainties. The distance fractional errors are below $10\%$ for $d < 5$ kpc, and below $20\%$ for $d_{kin} < 10$ kpc. The distances were estimated using a model designed to reproduce the properties of the data set in terms of spatial and luminosity distribution. $d_{kin}$ refers to the ‘astro-kinematic’ distances calculated by combining parallaxes and proper motions with a
 model for the expected velocity and density distribution of young stars as explained in  \cite{Zari2021}.

As previously mentioned, our focus is on young stars. Consequently, we have restricted the sample to stars younger than 500 Myr. Beyond 13 kpc, the reliability of our data significantly diminishes. Therefore, we have focused on stars within a Galactocentric radius range of 8 kpc to 13 kpc, which offers a sufficiently large sample size. To ensure data quality, we performed several filtering steps. We removed outliers with exceptionally large radial velocities ($>300, \mathrm{km/s}$), large errors in radial velocity ($>60, \mathrm{km/s}$), and high vertical actions ($J_z > 20, \mathrm{kpc , km/s}$), leaving about $\sim 7900$ OBA stars. The resulting filtered sample will be referred to as Z21-500 in subsequent sections (i.e., a clean version of the sample of \cite{Zari2021} with ages less than 500 Myr).

\begin{figure}
\centering
\includegraphics[width=0.5\columnwidth]{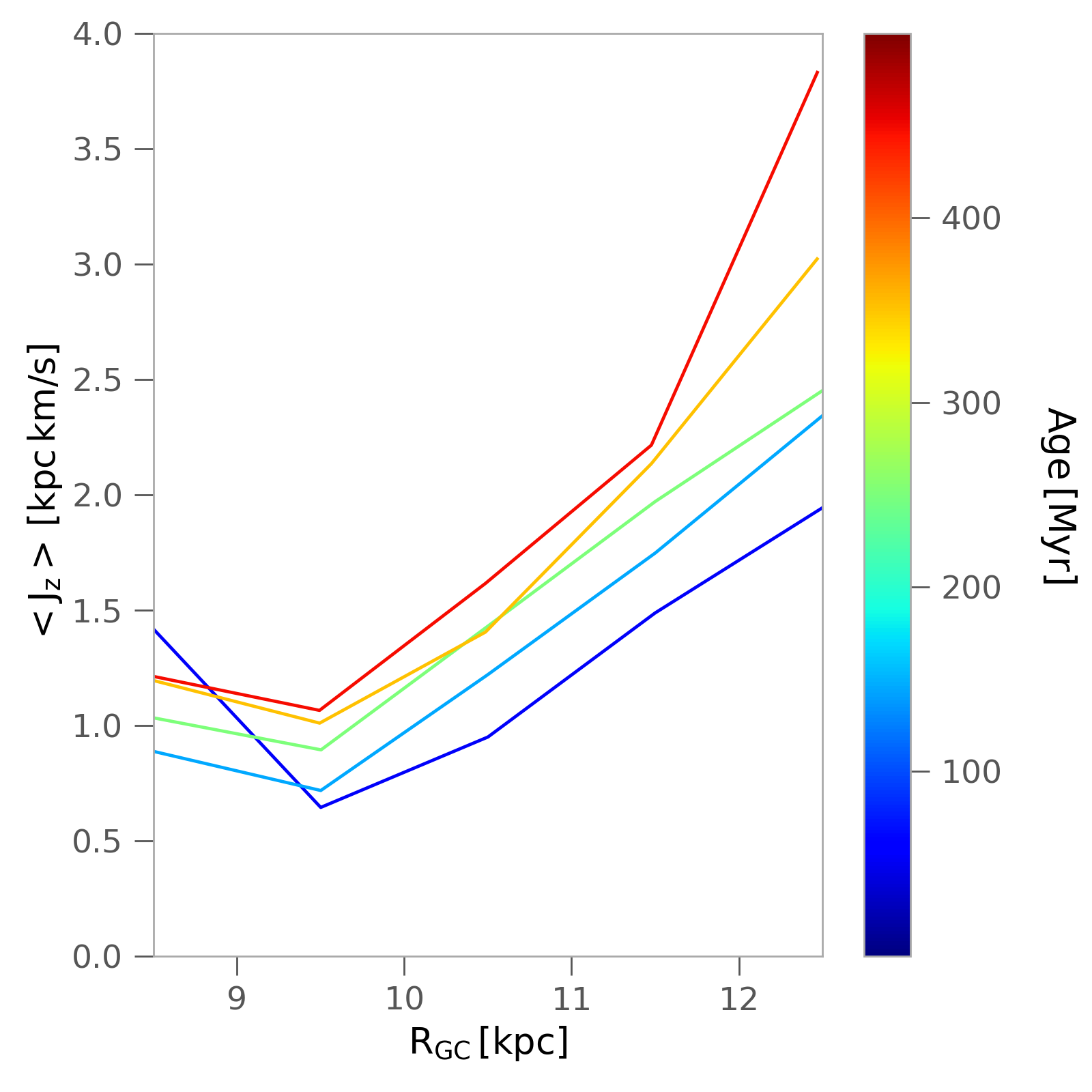}
\caption{Mean vertical action ${J}_{z}$ as a function of radius at different ages. The data were divided in bins of 100 Myr, each reflected by a differently colored line.}
\label{fig/model&data_jz_r}
\end{figure}

The computed phase-space coordinates enable us to visualize the spatial distribution of our star sample. In Figure \ref{fig:combined_distribution}, we show in the left panel the spatial distribution of stars in our sample Z21-500 in the $R - Z$ plane and in the right panel the spatial distribution of stars in our sample in the $X - Y$ plane. The latter was calculated using a kernel density estimator (KDE) with a Gaussian kernel and a bandwidth of $0.1$ kpc. The observed low-density gap at the center of the map (around $x, y \approx 8, 0$ kpc) may stem from the magnitude limit of LAMOST observations at the bright end, the intrinsically low population of OBA stars in the solar neighborhood, or the Sun's location in the Local Bubble \citep[see][]{Zari2021}. Without a comprehensive understanding of the joint selection function of the surveys, it is challenging to ascertain whether the observed overdensities and gaps represent artificial features or actual physical structures. The pencil-beam overdensities emanating from the Sun's position are indicative of the LAMOST survey's footprint.

%



To calculate the vertical actions $J_z$:
\begin{equation}
    J_z = \frac{1}{2\pi}\oint v_z dz.
\end{equation}
and radial actions $J_R$ of the stars in the sample, we use the axisymmetric Milky Way potential \texttt{MWPotential2014} \citep{bovy_2015_galpy}, scaled to have a circular velocity of 236 km$,\mathrm{s}^{-1}$, and the Galpy package's implementation of the Staeckel Fudge \citep{binney_2012}.
We binned the data by Galactocentric radius, using a bin size of 1 kpc and spanning a range from $R_{\mathrm{GC}} = 8$ to $13
$ kpc.  The distribution of $R$, ages, $J_R$, and $J_z$ is shown in Figure \ref{fig/hst_ages.png}. Most stars are centered close to $10$ kpc, with vertical and radial actions close to $
0$. Figure \ref{fig/model&data_jz_r} displays the vertical action as a function of Galactocentric radii, colored by different ages. $J_z$ increases with age, suggesting that young stars are born on cold orbits, i.e., with low vertical actions, and undergo subsequent heating after birth. Motivated by these findings, we aim to  analyze the initial actions of stars at birth and their heating over time.


%
%
%

%
%
%


\begin{figure*}
\includegraphics[width=\textwidth]{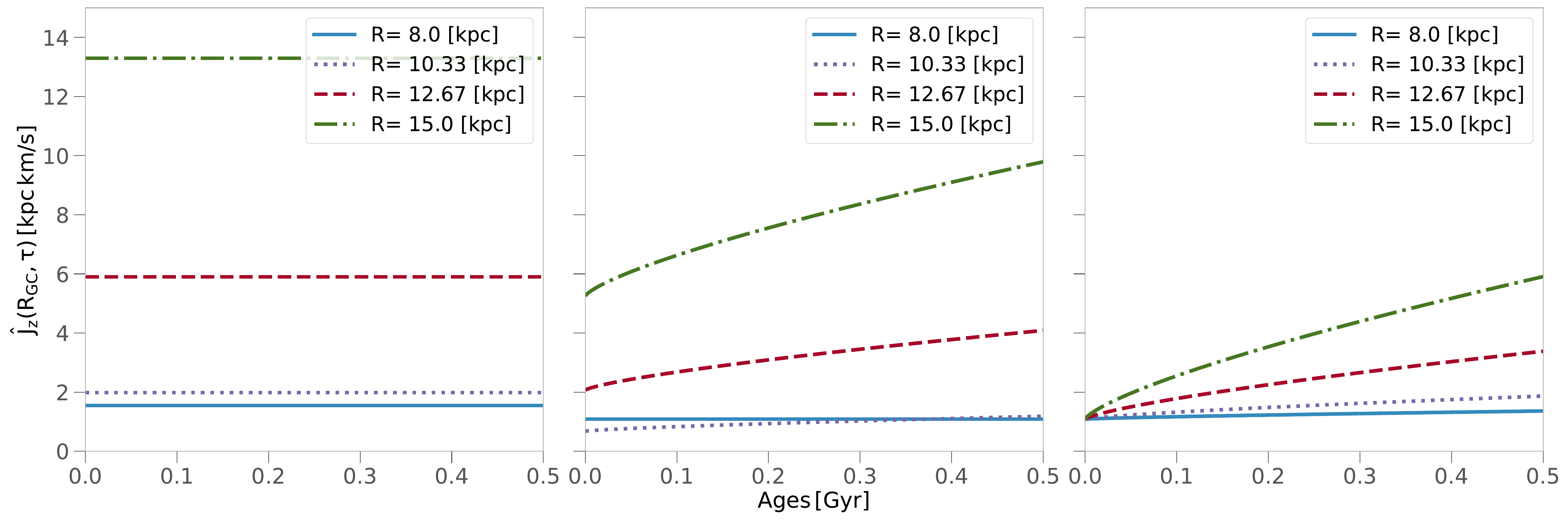}
\caption{Model presented in Eq.~\ref{eq:math_model} for three choices of model parameters, illustrating the various scenarios that the model can capture. The first panel illustrates a scenario where $\gamma =0$, which implies an increase in $\widehat{J}_z$ after birth does not scale with age. The middle panel shows a second scenario where $d=0$ (the simplified model), where the increase of $\widehat{J}_z$ in the last Gyr has no dependence on the radius. The third panel illustrates the third scenario $a=0, b=0$, i.e. where the stars' characteristic $\widehat{J}_z$ at birth does not depend on the Galactocentric radius.}
\label{fig/theoretical-model}
\end{figure*}

\section{Model of Vertical Action vs. Age and Galactocentric Radius}\label{sec:Vertical Heating Model }
In this section, we construct a physically motivated model to describe the vertical action at birth as a function of the Galactocentric radius and the subsequent temporal evolution of $J_z$ as a function of stellar age and radii\footnote{We could have equally well chosen to work with angular momentum rather than Galactocentric radius, but they are almost equivalent for young stars on near-circular orbits.}.

The model is then quantitatively compared to the data, taking into account observational uncertainties.

\subsection{Parameterized Model}\label{sec:Parameterized Model}
 
Due to the stochastic nature of the Galaxy, a deterministic model for the vertical actions of stars is not possible. However, we can describe with some probabilistic distribution the vertical actions of stars as a function of radius and age $p(J_z| R_{GC}, \tau)$. 

We start with the assumption of an isothermal distribution \citep{spitzer1942dynamics} in the harmonic limit \citep{binney_2010, binneyMacmillan2011models}.  This essentially means that there are many stars with small vertical actions and fewer stars with large vertical actions.
\begin{equation}
p\left(J_{z}\right) \sim \exp \left(-\nu J_{z} / \sigma_{z}^{2}\right).
\end{equation}
where, $\nu$ is frequency of vertical oscillations and $\sigma_{z}^{2}$ is the variance of vertical actions.

Due to the properties of the exponential distribution, we know that the mean of the vertical action is $\widehat{J}_{z}=\sigma_{z}^{2} / \nu$. Then a normalized distribution of $J_{z}$ can be written as

\begin{equation}
p\left(J_{z} \mid R_{\mathrm{GC}}, \tau\right) =\frac{1}{\widehat{J}_{z}\left(R_{\mathrm{GC}}, \tau\right)} \exp \left(-\frac{J_{z}}{\widehat{J}_{z}\left(R_{\mathrm{GC}}, \tau\right)}\right) .
\end{equation}

We model  the global relation of vertical action with age by assuming that the mean $J_z$ increases as a function of age as follows: 

\begin{equation}\label{eq:math_model}
    \widehat{J}_{z}\left(R_{\mathrm{GC}}, \tau\right)= \widehat{J}_{z, 0}\left(R_{\mathrm{GC}}\right) + \Delta \widehat{J}_{z, 1}\left(R_{\mathrm{GC}}\right) \left(\frac{\tau}{1 \mathrm{Gyr}}\right)^{\gamma}.
\end{equation}
Here, $\widehat{J}_{z, 0}\left(R_{\mathrm{GC}}\right)$ is the $J_{z}$ of stars at birth, and we assume that it is constant during the $500$ Myr studied here. $\widehat{\Delta J}_{z,1}$  is the typical increase in $J_{z}$ in the last Gyr; $\gamma$ describes the scaling of the heating with age, and we assume it can be described as a power law. Then, putting the Sun as the center, i.e, $ \Delta R_{\mathrm{GC}}=R_{\mathrm{GC}}-8 \mathrm{ kpc}$, we can write the global model 
as 
\begin{align}
\label{eq:jzo}
\widehat{J}_{z, 0}\left(R_{\mathrm{GC}}\right)&\equiv a  \Delta R_{\mathrm{GC}}^2+b \Delta R_{\mathrm{GC}} + c,\\
\Delta \widehat{J}_{z, 1}\left(R_{\mathrm{GC}}\right) &\equiv d  \Delta R_{\mathrm{GC}}^2+ e,\label{eq:jz1}
\end{align}
where the parameters are $\Theta=\{a,b,c,d,e\}$. The flexibility of the model and its capability of describing several scenarios is shown in Figure \ref{fig/theoretical-model} . The figure demonstrates three unique scenarios, effectively highlighting the model's capability to discern whether there is a correlation between vertical action and age, or a lack thereof.


The observed vertical action distribution can be written as

\begin{equation}
\begin{split}
    p(&J_{z,\text{obs}} \mid R_{\mathrm{GC}}, \star, \Theta) = \iint p\left(J_{z,\text{obs}} \mid J_{z,\text{true}}, \Theta \right) p\left(\tau_{\text{true}} \mid *\right) \\
    &\times p\left(J_{z,\text{true}} \mid R_{\mathrm{GC, true}}, \tau_{\text{true}}, \Theta\right) \, d\tau_{\text{true}} \, dJ_{z, \text{true}},
\end{split} \label{eq:integral}   
\end{equation}

where $*$ are the stellar parameters used to produce the age posteriors as described in section \ref{sec:sample}.

 We generate a \textit{noise model}, which refers to the process of accounting for uncertainties in the observed values, $x_{\text{obs}}$, for each star in the Z21-500 sample. Specifically, $x_{\text{obs}} = \{\mu_{\alpha, \text{obs}}, \mu_{\delta, \text{obs}}, \varpi_{\text{obs}}, \tau_{\text{obs}}\}$ represents the observed values, each subject to uncertainties modeled by $x_{\text{obs}} \sim \mathcal{N}(x,\sigma)$. Using a Gaussian distribution, we generate a set, or `cloud', of $N$ potential values for each observed star from that noise model. Then, we use this sample to approximate the integral in \ref{eq:integral}  numerically with importance sampling  as\\ $\sum_{i=1}^{N}p(J_{z,\text{true},i} \mid R_{\mathrm{GC, true}}, \tau_{\text{true}}, \Theta)$. The term in the sum effectively encapsulates the other terms in the integral because the sampling process (which generates \( J_{z,\text{true},i} \)) already accounts for the uncertainties in \( J_{z,\text{obs}} \) (via the noise model). Here, $N$ denotes the number of elements generated for each star within the noise model.

%
%
%
%
%
%
%
\subsection{Data Likelihood and Model Posteriors}\label{sec:Data Likelihood and Model Posteriors}

Since the data points are independent from each other, we can write the likelihood function as the product
\begin{equation}
    \mathcal{L}(p\left(\{J_{z,obs}\} \mid \{{R}_{\mathrm{GC}}\}, \Theta\right))=\prod_{i}^{N_s} p\left(J_{z,i,obs} \mid {R}_{\mathrm{GC},i}, \Theta\right),
\end{equation}\label{eq:likelihood}
which is maximized when $\Theta$ corresponds to the true values. We optimize the log-likelihood using SciPy's \citep{2020NatMe..17..261V} implementation of the Nelder-Mead algorithm to obtain the initial guess for the MCMC implementation.

We sample $\Theta$ from the posterior distribution using the Affine-invariant MCMC implementation \texttt{ emcee} \citep{foreman-mackey_2013}. We run \texttt{emcee} with 500 walkers for 10,000 steps per walker, discarding the first 500 steps. We fit the global model with the functional form in equations \ref{eq:math_model}, \ref{eq:jzo}, and \ref{eq:jz1}.


%
%
%
%
%
%
\section{ Results }\label{sec:Results}

\begin{figure*}
\includegraphics[width=\textwidth]{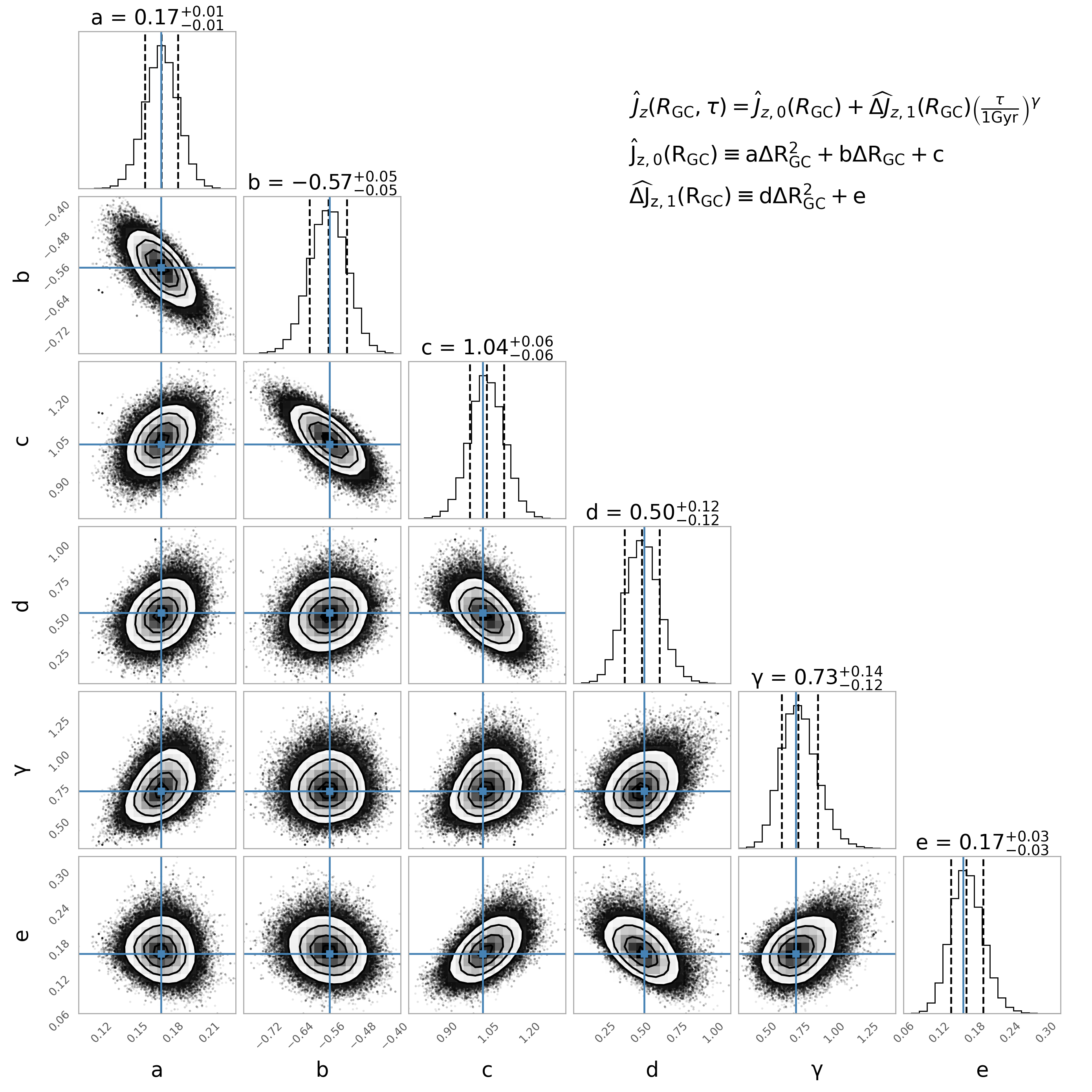}
\caption{Posterior distribution and two-parameter correlations for the model parameters, indicating there are a few covariances between some model parameters, but no strong degeneracy.}
\label{fig/corner_plot}
\end{figure*}

In this section, we summarize the results of our Bayesian parameter inference. Figure \ref{fig/corner_plot} presents a detailed overview of the posterior distributions and the correlations between parameters. Crosshairs in the figure mark the best-fit values, which are defined as the means of the marginalized distributions. We summarize these best-fit parameters in Table \ref{table:Best-fit}. The model parameters are highly non-degenerate and the MCMC chain converges to the best-fit model. 

\begin{figure}
\centering
\includegraphics[width=0.9\columnwidth]{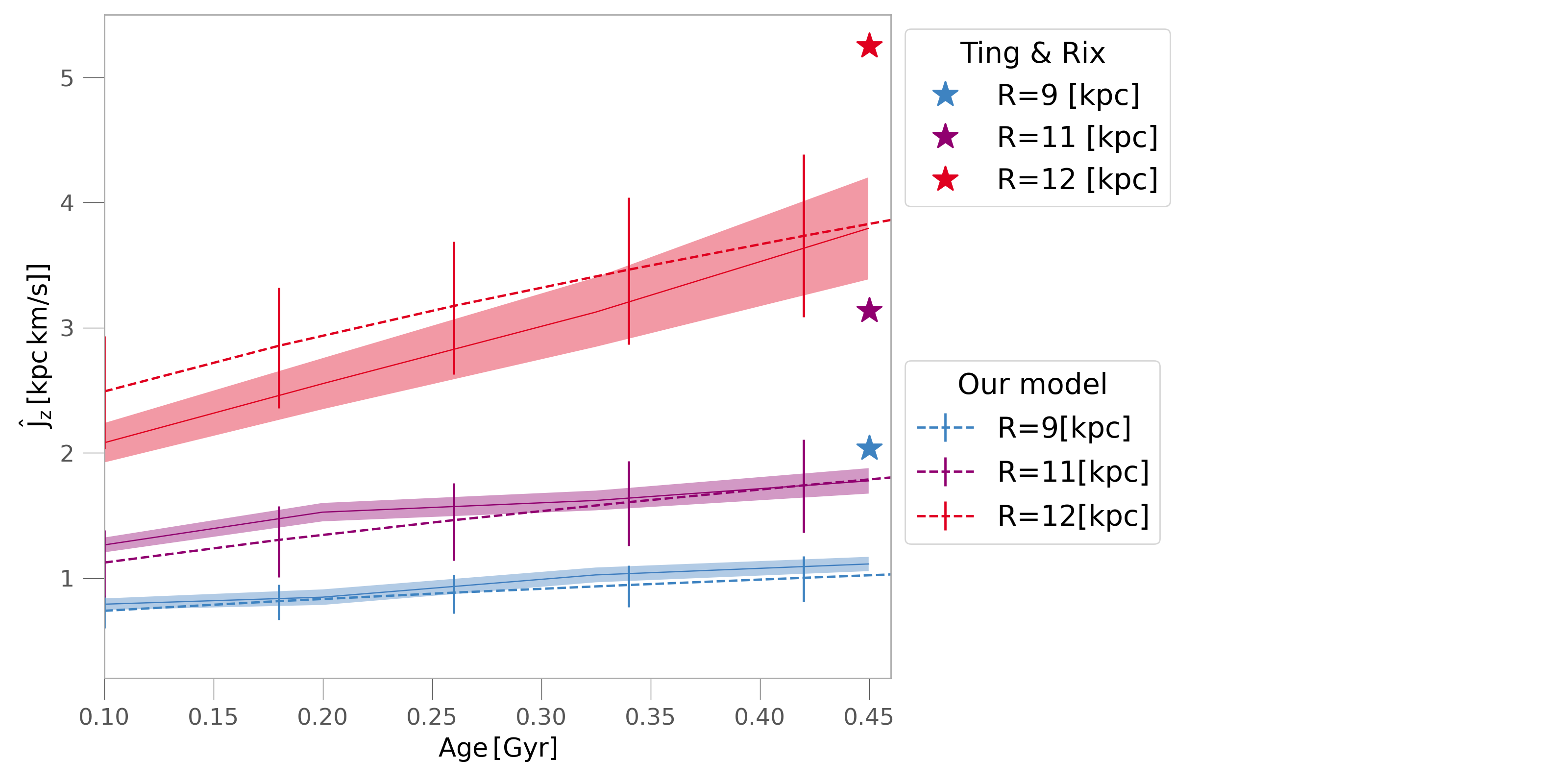}
\caption{ $\langle J_{z} \rangle$ as a function of age at different Galactocentric radii for the data (solid lines) and the best-fit model (dashed).  
The standard deviation is shown as shadows after performing  bootstrap analysis  to estimate the uncertainties. For comparison, we have included the results from \cite{ting_2019} as stellar marks at 0.45 Gyr. Their results are larger than ours by a factor $\sim 1.4$.
}
\label{fig/model&data_jz_age}
\end{figure}

\begin{table}[ht!]
\centering
\caption{The 16-50-84$^{th}$ percentiles of the marginalized posterior.}
\label{table:Best-fit}
\begin{tabular}{ccc}

\hline\hline
Parameter & Best Fit & Uncertainty \\ 
\hline
a & 0.17 & $^{+0.01}_{-0.01}$ \\ 
b & -0.57 & $^{+0.05}_{-0.05}$ \\
c & 1.04 & $^{+0.06}_{-0.06}$ \\
d & 0.50 & $^{+0.12}_{-0.12}$ \\
$\gamma$ & 0.73 & $^{+0.14}_{-0.12}$ \\
e & 0.17 & $^{+0.03}_{-0.03}$ \\
\hline
\end{tabular}
\end{table}


Figure \ref{fig/model&data_jz_age} 
illustrates the comparison between the observed data and the best-fit model predictions (with parameters listed in Table \ref{table:Best-fit}, derived from Eq. \ref{eq:math_model}). We bin the 
data in Galactocentric radius (9, 11, and 12 kpc) and age intervals (0.1 to 0.5 Gyr).  The shaded areas represent the standard deviation of the observed data. This was calculated using a bootstrap analysis with 500 samples, where each sample was generated by resampling the original data with replacement. For each bootstrap sample, we calculated the mean and median $J_z$ values within defined radial and age bins. We use the MCMC chains to get the uncertainties on the model, depicted as vertical lines around the dashed lines. 

The vertical action $p\left(J_{z} \mid {R}_{\mathrm{GC}}\right)$ increases linearly with age.
Even though we assumed stars are born cold, they show non-zero vertical action at birth.
The vertical heating rate increases at large $R_{GC}$ (Figure \ref{fig/model&data_jz_age}). 

\begin{figure*}
\centering
\includegraphics[width=1.05\linewidth]{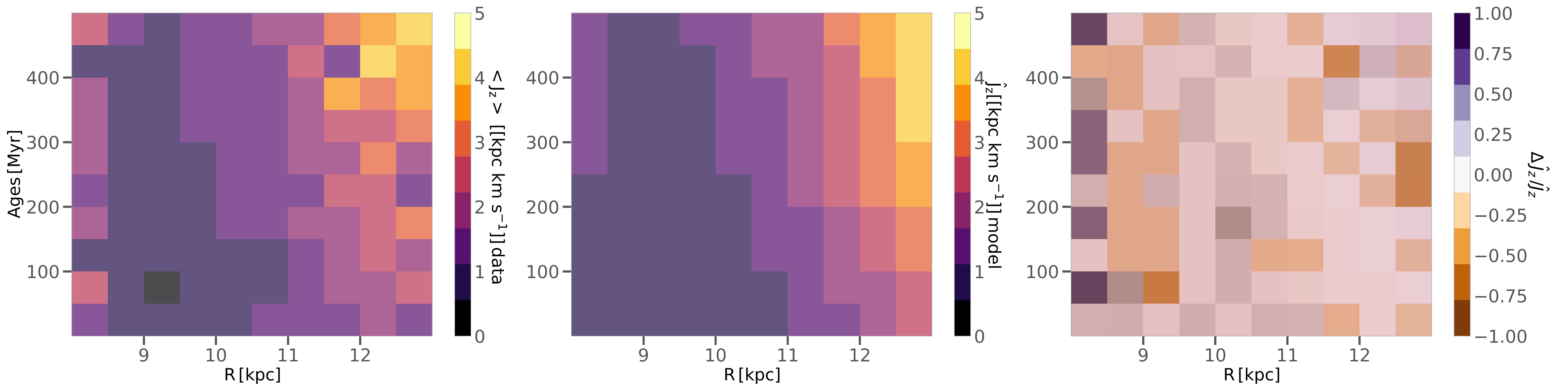}
\caption{Mean vertical actions in the age-Galactocentric radius plane: The first panel is colored by the mean $J_z$ from the data. The second panel is colored by $\widehat{J}{z}(R{\mathrm{GC}})$, as obtained from the model. The third panel shows the difference in $J_z$ between the data and the model.}
\label{age_r_jzcolored_horizontal_residual}
\end{figure*}

The left panel of figure \ref{age_r_jzcolored_horizontal_residual} shows the values of $J_{z}$ calculated from the data. The central panel shows the median $\widehat{J}_{z}$ calculated by the model with the best-fit parameters shown in Table \ref{table:Best-fit}. Both panels show a gradient in the diagonal direction as $J_{z}$ increases with respect to the ages and Galactocentric radii. The right panel shows the difference between the model and the data, normalized with respect to the value of the model $\widehat{J}_{z}$.

\section{Discussion and Conclusion}\label{sec:Discussion and Summary}

Using a substantial data set of young massive stars drawn from \cite{Zari2021} and cross-matched with LAMOST to obtain radial velocities and ages, we measured the vertical motions of young stars ($<500$Myr) in the Galactic disk over an unprecedentedly wide range in Galactocentric radii (8 - 13~kpc).  
Previous studies over such a wide range in $R_{\mathrm{GC}}$ were based on red clump stars \citep{ting_rix_2019}, which represent older populations (1.5-10~Gyr). Our comparisons with \cite{ting_rix_2019} (see Figure \ref{fig/model&data_jz_age}) are not directly equivalent due to methodological differences. Specifically, our model does not account for selection effects, and neither our model nor the \cite{ting_rix_2019} observations account for dust extinction. This is the primary reason we initially did not consider the potential flaring in our analysis; instead, our focus was on the evolutionary trends in $J_z$ rather than absolute values. Note that we fit only one model with functional radial dependence rather than radial bins, so any rigidity of the model might result in a small offset between the mean of the data and the best-fit model, as can be seen for $R=12$~kpc. Overall, the agreement between our model predictions and our observed data is satisfactory; see Figure~\ref{age_r_jzcolored_horizontal_residual}. Notice, the model of \cite{ting_rix_2019} is also shown as star points and differs by a factor $\sim 1.4$ ; this difference could result from using different tracers or a different survey selection.

Our analysis centered on characterizing vertical motions through vertical action ($J_z$), fitting the data with a model that predicts the mean vertical action $\widehat{J}_{z}(\tau~|~R_{GC})$.  We found that (1) the stars' mean vertical action $\widehat
{J}_{z}(\tau~|~R_{GC})$ increases nearly linearly with age at all $R_{GC}$, and (2)  that $\widehat{J}_{z}(\tau~\mid ~R_{GC})$ increases towards larger Galactocentric radii for any given age, including at birth. Here, we interpret only the former: the age-dependence of the mean vertical action, but not the latter, which can arguably be tied to extinction decreasing our probability to observe young stars in the midplane or to the Galactic warp.

To effectively model the observed relationship between mean vertical action with respect to age and Galactocentric radius, we explored several models, starting with a simplified approach that included only three terms: a constant term, a linear term with respect to radius, and a quadratic term with respect to radius. Subsequently, we extended our investigation to more intricate models incorporating quadratic dependencies on age.  Among the range of models explored, the model presented in this paper emerged as the optimal choice, giving the most robust and consistent fit to the data.

As stated in earlier sections (see Section \ref{sec:sample}), we have chosen a fixed uncertainty value of $10\,\text{km/s}$ for the line-of-sight velocities ($v_\mathrm{los}$) of all the stars in our analysis. Nevertheless, as part of our investigation, we conducted a test by incorporating the uncertainties provided by LAMOST into the MCMC analysis and found the same best fit parameters within 1 sigma.

It is tempting and traditional to interpret the observed increase in vertical action with age in terms of \emph{heating} because the near-linear age dependence of the mean vertical action ($\gamma = 0.73\approx 1$) is consistent with heating by giant molecular clouds and with observations of older stellar populations \citep{ting_rix_2019}.
Such vertical heating could be driven by a combination of giant molecular clouds, external perturbations such as satellites and flybys, or by internal asymmetries. 
However, internal heating mechanisms do not appear to be likely explanations of the data, or at least not uniquely, because the data require that such heating would need a heating rate constant with $R_{GC}$, even though the density of potentially scattering molecular clouds should drop off with $R_{GC}$ \citep[e.g.,][]{rice2016uniform}.

There may be alternative explanations for our observed $\widehat{J_z}(\tau~|~R_{GC})$. We discuss them below; however, exploring these would necessitate a careful, in-depth analysis and model selection that extends beyond the  scope of this work. For instance, stars at increasingly larger radii may originate from gas that is not co-planar with the inner Galactic disk, such as in a warped gas disk \citep[refer to Figure 8 in][]{Soler2022}.
Alternatively, the mean increase in vertical action with age could be attributed to the azimuthal mixing of stars over time. Azimuthal mixing refers to the redistribution of stars along their orbital paths around the Galactic center due to differential rotation. If a warp is present in the disk, the mid-plane is not centered on $z=0$ everywhere, but it is a function of radius and azimuth. Stars born in this warped region will have initial vertical positions ($z$) that are not zero, and therefore they will have computed actions $J_z$ that are not zero. Their angular velocities decrease with increasing radius ($\Omega_{\phi} \propto 1/R$). Consequently, stars in the outer disk take longer to reach the solar neighborhood. So if there is a warp, then the vertical action at birth ($J_z$) is also a function of azimuth ($\phi$). Even in the absence of heating, if we select stars in the solar neighborhood, there will be a relation between their age and their birth azimuth ($\phi = \phi_0 + \Omega_\phi t$), which translates into a relation between $J_z$ and age ($t$). Pursuing further explanation would require modeling or simulation analysis, which lies beyond the scope of this paper.
 Should there be a trend of $J_z$ with azimuth at birth, subsequent azimuthal mixing could manifest as an apparent age trend within a localized sample. Pursuing this explanation would require modeling or simulation analysis, which lies beyond the scope of this paper.

Massive stars are often found in binary systems \citep{Sana2014ApJ}. Consequently, some stars in our sample could be part of such systems, potentially influencing the kinematics. However, binarity does not seem to be a plausible source of the observed trend of $J_z$. There is no evidence to suggest that the impact of binarity escalates with advancing age or at greater Galactocentric radii.
Similarly, after massive binaries form, one of the companions may explode in a supernova before the other, thereby unbinding from its companion and leaving it walking (or running) at the speed it had when they were bound. 
Typical ejection velocities range from 5-20 km$,\mathrm{s}^{-1}$ and exhibit only a weak dependence on the mass (and thus, the age) of the ejected star \citep{renzo2019}. Thus, this probably not be the origin of our observed $J_z - \tau$ trend. 

Our results are likely caused by a combination of factors, including vertical heating by giant molecular clouds and the structural evolution of the Galactic disk. There is a complex interplay of multiple mechanisms and this complexity warrants a cautious interpretation. In this study, we have not considered the potential under-representation of mid-plane stars in our sample due to interstellar extinction or effects of LAMOST selection function. Figure \ref{fig:combined_distribution} shows some lack of data around $z \approx 0$ and the decrease in the number of stars beyond 11 kpc.
 Qualitatively, interstellar extinction, leading to an increasing scarcity of mid-plane stars with distance from us, would result in an increase of the mean vertical action with distance from the Sun (and thus, in this case, towards the outer disk). Consequently, without modeling of dust and selection functions, the radial dependence of mean vertical motion should be interpreted with caution. However, using a toy model, we confirmed that our main finding—that the mean vertical action increases nearly linearly with age across all Galactocentric radii—remains robust. We produced a toy disk model, generating stellar 3D positions and velocities, and where the mean vertical action increases with age. Then, we applied a toy dust model where stars with galactic latitudes larger than some threshold do not make it into the mock dataset. From this experiment, we found (1) a non-physical increase in $\langle J_z \rangle (R)$ with radius, as expected and (2) no effect on the $\langle J_z \rangle (\tau)$ relation, which gives us confidence that our main result is robust.

Finally, we explored only a small volume near the Sun due to the reduced sky coverage of the sample, as determined by the footprint of the LAMOST survey.  The all-sky coverage of future spectroscopic surveys, such as SDSS-V \citep{sdssDR18, Kollmeier2017}, will allow to extend this to a larger volume of the Milky Way disc. Enhanced coverage of the disk in Galactic azimuth will facilitate distinguishing whether the observed trends arise from a combination of heating processes or intrinsic birth properties, such as disk asymmetries or warping.

\section{Acknowledgments}

This work was partially carried out during the MPIA Summer Internship, funded by the Max Planck Society.

This work has made use of data from the European Space Agency (ESA) mission
{\it Gaia} (\url{https://www.cosmos.esa.int/gaia}), processed by the {\it Gaia}
Data Processing and Analysis Consortium (DPAC,
\url{https://www.cosmos.esa.int/web/gaia/dpac/consortium}). Funding for the DPAC
has been provided by national institutions, in particular the institutions
participating in the {\it Gaia} Multilateral Agreement.

This work
has used data from the Guoshoujing Telescope, where the LAMOST
survey was executed. LAMOST is a National Major Scientific Project built by
the Chinese Academy of Sciences\citep{zhao2012lamost, deng2012lamost}.

NF acknowledges the support of the Natural Sciences and Engineering Research Council of Canada (NSERC), [funding reference number 568580] through a CITA postdoctoral fellowship and acknowledges partial support from an Arts \& Sciences Postdoctoral Fellowship at the University of Toronto.

This research made use of Astropy \citep{robitaille2013astropy, 2022ApJ...935..167A, 2018astropy, 2013Astropy}, matplotlib \citep{hunter2007matplotlib}, numpy \citep{harris2020array}, and SciPy \citep{2020NatMe..17..261V}. We extend our gratitude to the reviewer for their insights and constructive feedback.


\newpage
\bibliography{lit}

\end{document}